\documentclass[prl,twocolumn,superscriptaddress,showpacs,amsmath,amssymb,floatfix,amstex]{revtex4-1}

\usepackage{units}
\usepackage{psfrag}

\usepackage[dvips]{graphicx}

\begin{document}

\title{Dynamics of multifrequency oscillator communities}

\author{Maxim Komarov} 
\affiliation{\mbox{Department of Control Theory, Nizhni Novgorod University, 
  Gagarin Av. 23, 606950, Nizhni Novgorod, Russia}}
\affiliation{\mbox{Department of Physics and Astronomy, Potsdam University, 
  Karl-Liebknecht-Str 24, D-14476, Potsdam, Germany}}
\author{Arkady Pikovsky} 
\affiliation{\mbox{Department of Physics and Astronomy, Potsdam University, 
  Karl-Liebknecht-Str 24, D-14476, Potsdam, Germany}}

\date{\today}

\begin{abstract}
We consider a generalization of the Kuramoto model of coupled oscillators to the
situation where communities of oscillators having essentially different natural frequencies interact. General equations describing possible resonances between the communities' frequencies are derived. The mostly simple situation of three
resonantly interacting groups is analyzed in details. We find conditions for the mutual coupling to promote or suppress synchrony in individual populations, and present examples where interaction between communities leads to their synchrony, or to a partially asynchronous state, or to a chaotic dynamics of order parameters.

\end{abstract}


\maketitle
Networks of coupled oscillators describe synchronization in lasers~\cite{Antyukhov_et_al-86}
and Josephson junctions~\cite{Wiesenfeld-Swift-95}, atomic recoil lasers~\cite{Javaloyes-Perrin-Politi-08}, electrochemical oscillators~\cite{Kiss-Zhai-Hudson-02a}, applauding persons in a large audience~\cite{Neda_etal-00}, pedestrians on footbridges~\cite{Eckhardt_et_al-07} and many other systems. In the simplest setup, when all oscillators are of a same type and coupled via a mean field, the synchronization transition has been treated by Kuramoto~\cite{Kuramoto-75,Kuramoto-84} in analogy with the  mean field theory of ferromagnetic phase transitions. Since that the Kuramoto model, where the coupled oscillators are represented through the phase dynamics,
 has been used as a paradigmatic one for mutual synchronization of oscillators~\cite{Kuramoto-84,Pikovsky-Rosenblum-Kurths-01,Acebron-etal-05}. In various generalizations this approach has been 
 extended to the more complex and general situations. One direction is introduction of complex coupling functions~\cite{Hansel93}, with possible nonlinear dependencies on the mean fields~\cite{Rosenblum-Pikovsky-07,Filatrella-Pedersen-Wiesenfeld-07,Pikovsky-Rosenblum-09}. Another very popular extension of
 the Kuramoto model deals with heterogeneous oscillator populations, 
 consisting  of different communities (groups) that differ in their contributions to the mean fields and in their response to these fields~\cite{Barreto-Hunt-Ott-So-08,Abrams-Mirollo-Strogatz-Wiley-08,Pikovsky-Rosenblum-11,Anderson_etal-12}. In particular, nontrivial regimes appear if some interactions are ``attractive'' and other ``repulsive'', or the oscillators can be characterized as ``conformists'' and ``contrarians''~\cite{Hong-Strogatz-11}. 
 
 In most studies of the interacting oscillator communities it is assumed that all oscillators have close frequencies around some basic one. 
For a small coupling this allows one to obtain, by virtue of averaging over the basic period, equations containing phase differences only, and to apply the Kuramoto method to them. 
 In this letter we extend the theory to the case of \textit{multifrequency} communities, where natural frequencies of interacting groups differ significantly. In this case one cannot perform a common averaging, but has to check if resonances between different communities are present. In our previous study we have focused on the
 non-resonant case~\cite{Komarov-Pikovsky-11}, in this letter we show that \textit{resonant} interactions between communities lead to non-trivial effects of mutual synchronization and desynchronization of groups, and also to 
 chaotic behavior of the mean fields.

We start with a formulation of general phase equations for resonantly interacting
oscillator communities. Oscillators are described by their phases, and interact via mean fields, produced by communities. A field produced by a community with index $m$ can be thus represented as a function of the generalized order parameters~\cite{Daido-96} of this community 
\begin{equation}
Z_k^{(m)}=\langle e^{ik\varphi^{(m)}}\rangle
\label{eq:mfdef}
\end{equation}
where $\langle\rangle$ means averaging over the community members.
A combination of these fields constitutes a force $Q(t)$, acting on an oscillator from  community $0$,
having frequency close to $\omega_0$, which influences  its phase dynamics as~\cite{Kuramoto-84,Pikovsky-Rosenblum-Kurths-01}
\[
\dot\varphi=\omega_0+\Delta\omega+S(\varphi)Q(t)=\omega_0+\Delta\omega+\sum s_n e^{in\varphi}Q(t)
\]
where $S(\varphi)=\sum s_n e^{in\varphi}$ is the phase sensitivity function of the oscillator and $\Delta \omega$ is a small individual deviation from $\omega_0$.
Representing this phase as $\varphi=\omega_0 t+\tilde\varphi$, where $\tilde\varphi$ varies slowly on the time scale $\omega_0^{-1}$, we can average over the period $2\pi/\omega_0$ to get
\begin{equation}
\dot{\tilde\varphi}=\Delta\omega+\sum s_n e^{in\tilde\varphi}q_{n\omega_0}
\label{eq:pheqav}
\end{equation}
where $q_{n\omega_0}=\int_0^{2\pi/\omega_0}\;dt\;Q(t)\exp[in\omega_0 t]$ is the
component of the forcing at the frequency 
$-n\omega_0$. In this component we have 
to consider the slowly varying ingredients of the forcing as the ``frozen'' ones. To separate slow and fast time scales in the forcing, we represent the order parameters as  
\[
 Z_k^{(m)}=\tilde{Z}_k^{(m)}e^{ik\omega_m t}
 \]
where $\omega_m$ is a basic frequency of the 
community with index $m$, and $\tilde Z$ are slow. Substituting this in $Q$ and expanding in powers of order parameters, we can generally write 
\begin{equation}
\begin{aligned}
q_{n\omega_0}&=\sum_{k,m}\tilde{Z}_k^{(m)}\delta(k\omega_m-n\omega_0)+\\
&\sum_{k,m,l,j}\tilde{Z}_k^{(m)}\tilde{Z}_l^{(j)}\delta(k\omega_m+l\omega_j-n\omega_0)+\ldots
\end{aligned}
\label{eq:gcf}
\end{equation}
We see that a direct interaction between communities $m$ and $0$ is possible if their basic frequencies are in a rational relation $k\omega_m=n \omega_0$ (if this relation is fulfilled only approximately, one uses the freedom in the definition of the basic frequency of the community, and shifts $\omega_m$ slightly to have an exact resonance). Furthermore, the second term in (\ref{eq:gcf}) describes a ``triplet'' interaction if three communities have frequencies satisfying $k\omega_m+l\omega_j\approx n\omega_0$; also higher-order interactions between four communities, described by cubic in $\tilde Z$ terms, are possible, etc.

In this letter we do not aim to consider all possible cases of resonant multifrequency interactions contained in (\ref{eq:pheqav},\ref{eq:gcf}) but focus on a simple example. We assume the simplest situation where the phase sensitivity function has only the basic harmonics $n=\pm 1$, and only the first-order mean fields with $k=\pm 1$ in (\ref{eq:mfdef}) contribute to the coupling. In this case the basic resonant condition includes three communities:
$\omega_1+\omega_2=\omega_3-\Delta$, with small mismatch $\Delta$. Taking also into account interactions inside communities (which are described by the first term in (\ref{eq:gcf}) with $\omega_m=\omega_0$), we end up with the phase model describing the resonant interaction of oscillators in three communities (cf.~\cite{Lueck-Pikovsky-11} for a particular case of $1:2$ resonance). To simplify notations, we denote the phases in these communities as $\phi_k,\psi_k,\theta_k$, and the corresponding order parameters as $z_1=\langle e^{i\phi}\rangle,\;z_2=\langle e^{i\psi}\rangle,\;z_3=\langle e^{i\theta}\rangle$:
\begin{equation}
\begin{aligned}
\dot\phi_k&=\omega_1+\Delta\omega_{1,k}+2\text{Im}[(\epsilon_1 z_1 +\gamma_1 z_2^*z_3) e^{-i\phi_k}]\\
\dot\psi_k&=\omega_2+\Delta\omega_{2,k}+2\text{Im}[(\epsilon_2 z_2 +\gamma_2 z_1^*z_3) e^{-i\psi_k}]\\
\dot\theta_k&=\omega_3+\Delta\omega_{3,k}+2\text{Im}[(\epsilon_3 z_3 +\gamma_3 z_1z_2) e^{-i\theta_k}]
\end{aligned}
\label{eq:3int}
\end{equation}
Here terms $\Delta\omega_{1-3,k}$ account for a distribution of frequencies of individual oscillators within communities, $\epsilon_{i}=\varepsilon_i e^{i\alpha_i}$ and $\gamma_{i}=\Gamma_ie^{i\tilde\beta_i}$ are complex coupling constants. In the absence of mutual resonant coupling ($\Gamma=0$), each community is described by the standard Kuramoto-Sakaguchi model~\cite{Sakaguchi-Kuramoto-86}. It is instructive to write microscopic phase equations for the phases:
\begin{equation}
\begin{aligned}
\dot\phi_k&=\omega_1+\Delta\omega_{1,k}
+\varepsilon_1 \sum_{j}\sin(\phi_j-\phi_k+\alpha_1)+\\&
\Gamma_1 \sum_{m,l}\sin(\theta_m-\psi_l-\phi_k+\tilde\beta_1)\\
\dot\psi_k&=\omega_2+\Delta\omega_{2,k}
+\varepsilon_2 \sum_{j}\sin(\psi_j-\psi_k+\alpha_2)+\\&
\Gamma_2 \sum_{m,l}\sin(\theta_m-\phi_l-\psi_k+\tilde\beta_2)\\
\dot\theta_k&=\omega_3+\Delta\omega_{3,k}
+\varepsilon_3 \sum_{j}\sin(\theta_j-\theta_k+\alpha_3)+\\&
\Gamma_3 \sum_{m,l}\sin(\phi_m+\psi_l-\theta_k+\tilde\beta_3)
\end{aligned}
\label{eq:3intph}
\end{equation}

To obtain a closed system of equations for the order parameters $z_{1,2,3}$, we adopt the Ott-Antonsen approach~\cite{Ott-Antonsen-08,Ott-Antonsen-09}, in which a particular form of the distribution of the phases is assumed, parameterized by the order parameter. If, furthermore, a Lorentzian distribution of frequencies around the basic ones is considered (with widths $\delta_{1,2,3}$), the Ott-Antonsen equations take especially simple form:
\begin{equation}
\begin{aligned}
\dot z_1&=z_1(i\omega_1-\delta_1) +(\epsilon_1z_1+\gamma_1 z_2^*z_3 -z_1^2(\epsilon_1^* z_1^*+\gamma_1^* z_2z_3^*))\\
\dot z_2&=z_2(i\omega_2-\delta_2) +(\epsilon_2z_2+\gamma_2 z_1^*z_3 -z_2^2(\epsilon_2^* z_2^*+\gamma_2^* z_1z_3^*))\\
\dot z_3&=z_3(i\omega_3-\delta_3) +(\epsilon_3z_3+\gamma_3 z_1z_2 -z_3^2(\epsilon_3^* z_3^*+\gamma_3^* z_1^*z_2^*))
\end{aligned}
\label{eq:3op}
\end{equation}
System (\ref{eq:3op}), describing the dynamics of order parameters of three resonantly interacting communities, is the main object of our analysis below. We focus on specific features resulting from the mutual coupling, where it acts ``against'' the internal coupling within the communities. 

\begin{figure}
\centering
\includegraphics[width=0.7\columnwidth]{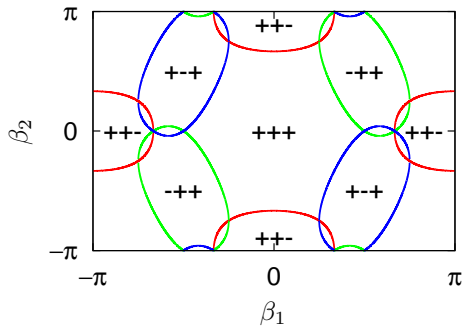}
\caption{(Color online) Regions of synchronizing and desynchronizing effect from the triplet coupling, for different combinations of arguments of coupling constants $\beta_1,\beta_2$. The effect on a community is marked by $(+)$ for enhancing synchrony, and by $(-)$ for a desynchronizing action. In small overlaps of the ovals these markers ``multiply'' (so, e.g. in the overlap of ovals $(+-+)$ and $(++-)$ we have
$(+--)$, i.e. community 1 synchronizes by coupling while communities 2,3 desynchronize).  }
\label{fig:stdiag}
\end{figure}

Essential properties of the coupling between oscillators such as their tendency to synchrony or to asynchrony, depend on the arguments of the complex coupling parameters $\alpha_i$ and $\tilde\beta_i$. For the coupling inside a community, the argument $\alpha$ corresponds to the phase shift in the oscillator-to-oscillator coupling in the Kuramoto-Sakaguchi formulation~\cite{Sakaguchi-Kuramoto-86}, for $\cos\alpha>0$ the coupling is attracting and synchronizing, while for $\cos\alpha<0$ it is repulsing and desynchronizing. A corresponding interpretation of arguments of mutual coupling $\tilde\beta_i$ is not so straightforward. To achieve it, we rewrite
the complex system (\ref{eq:3op}) in terms of the amplitudes and the phases of the complex order parameters $z_i=\rho_i\exp[i\Phi_i]$:
\begin{equation}
\begin{aligned}
\dot\rho_1&=-\delta_1\rho_1+(1-\rho_1^2)
(\varepsilon_1\rho_1\cos\alpha_1+\Gamma_1\rho_2\rho_3\cos(\Psi+\beta_1))\\
\dot\rho_2&=-\delta_2\rho_2+(1-\rho_2^2)
(\varepsilon_2\rho_2\cos\alpha_2+\Gamma_2\rho_1\rho_3\cos(\Psi-\beta_1))\\
\dot\rho_3&=-\delta_3\rho_3+(1-\rho_3^2)
(\varepsilon_3\rho_3\cos\alpha_3+\Gamma_3\rho_1\rho_2\cos(\Psi-\beta_2))\\
\dot\Psi&=\Delta'-(\rho_3^{-1}+\rho_3)\Gamma_3\rho_1\rho_2\sin(\Psi-\beta_2)-\\&
(\rho_2^{-1}+\rho_2)\Gamma_2\rho_1\rho_3\sin(\Psi+\beta_1)-\\&
(\rho_1^{-1}+\rho_1)\Gamma_1\rho_2\rho_3\sin(\Psi-\beta_1)
\end{aligned}
\label{eq:3opreal}
\end{equation}
Here $\Delta'=\Delta+(1+\rho_3^2)\varepsilon_3\sin\alpha_3-(1+\rho_2^2)
\varepsilon_2\sin\alpha_2-(1+\rho_1^2)
\varepsilon_1\sin\alpha_1$ is the effective frequency mismatch that includes also frequency shifts due to intra-communities interactions,
 $\Psi=\Phi_3-\Phi_2-\Phi_1+0.5(\tilde\beta_1+\tilde\beta_2)$ is the phase difference between the communities order parameters, $\beta_1=0.5(\tilde\beta_1-\tilde\beta_2)$ and $\beta_2=\tilde \beta_3+0.5(\tilde\beta_1-\tilde\beta_2)$. One can see from the equations for $\dot\rho_i$, that the effect of the inter-community coupling depends on signs 
of terms $\cos(\Psi\pm\beta_1),\cos(\Psi-\beta_2)$. These depend on the dynamics of the phase difference $\Psi$, so a general conclusion is hardly possible. Let us look on the simplest situation of exact resonance, where $\Delta'=0$, of
equal coupling constants $\Gamma_1=\Gamma_2=\Gamma_3$, and of equal order parameters in each community $\rho_1=\rho_2=\rho_3$. Then the stable phase difference is $\Psi_0=\arctan(\sin\beta_2(\cos\beta_2+2\cos\beta_1)^{-1})$. 
Substituting this solution, we come to the diagram Fig.~\ref{fig:stdiag}, which shows the regions of positive and negative signs of factors $\cos(\Psi_0\pm\beta_1),\cos(\Psi_0-\beta_2)$, i.e. the regions where mutual resonance coupling between communities promotes (for positive signs) synchrony
or tends to desynchronize (for negative signs). In analytical form, these conditions are: Communities $1,2,3$ synchronize for $2\cos^2\beta_1+\cos(\beta_2+\beta_1)>0$, $2\cos^2\beta_1+\cos(\beta_2-\beta_1)>0$, and $1+2\cos\beta_1\cos\beta_2>0$, respectively, and desynchronize otherwise. We see that desynchronization in all three communities is not possible, while there are situations where two desynchronize, regimes where one desynchronizes, and states
where mutual interaction improves synchrony in all communities. Below we give examples for synchronization and desynchronization effects in several setups.

\begin{figure}
\centering
(a)\includegraphics[width=0.7\columnwidth]{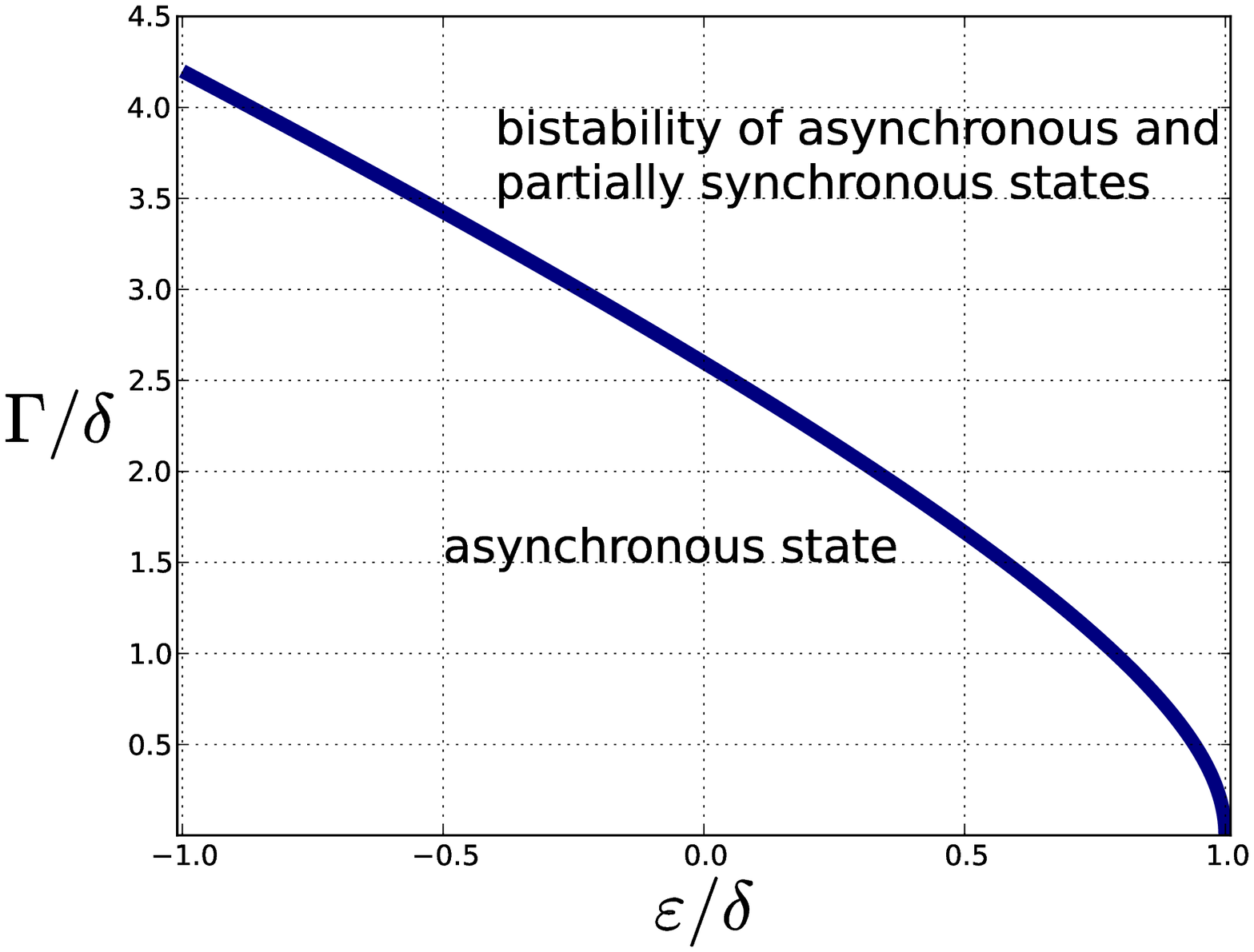}
(b)\includegraphics[width=0.7\columnwidth]{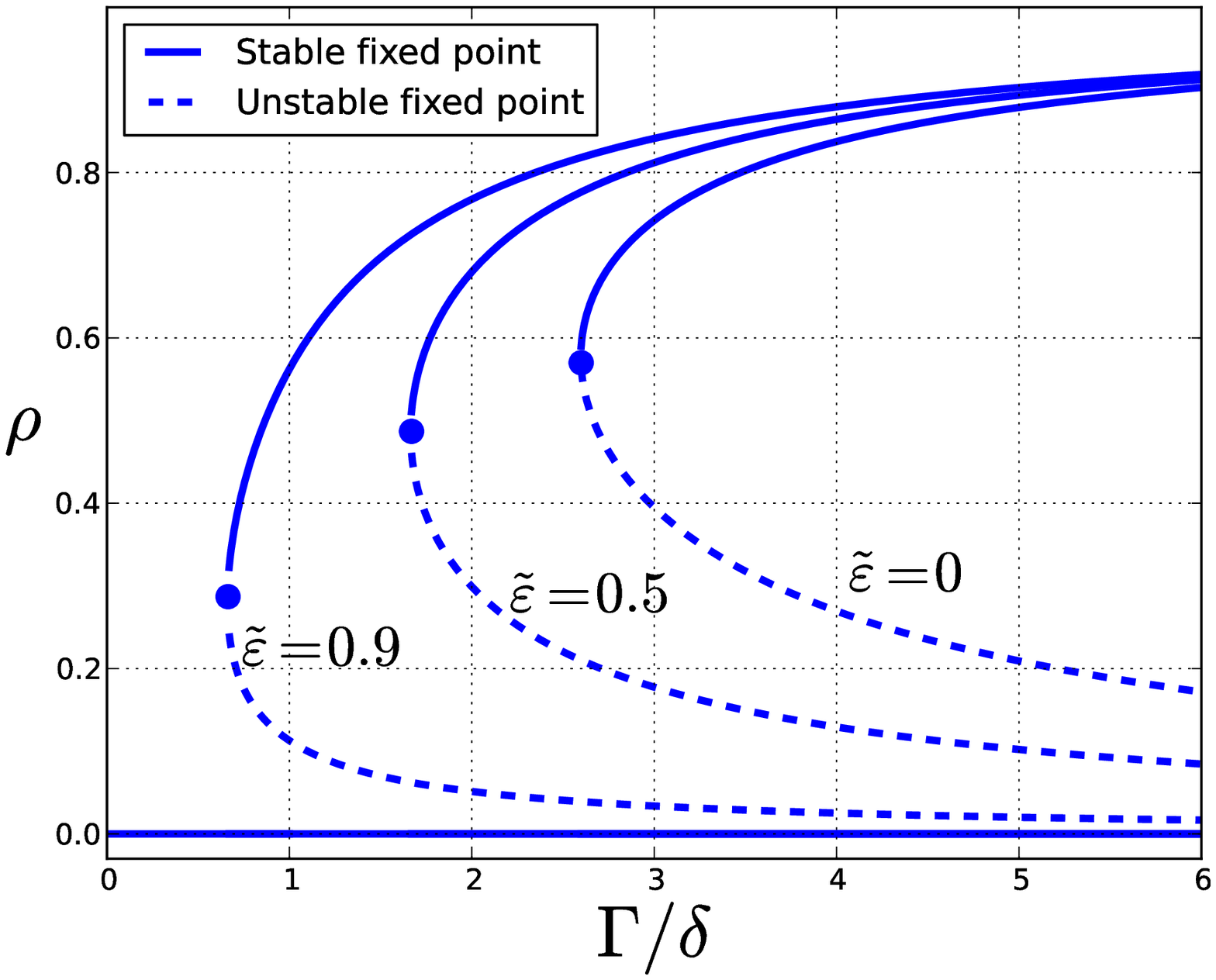}
\caption{(Color online)(a) Region on the plane of parameters where partially synchronous state appears. (b) Bifurcation diagrams showing dependence of the steady state order parameters (equal for all communities) on the mutual coupling, for different couplings $\tilde\varepsilon=\varepsilon/\delta$ within the groups.}
\label{fig:msyn}
\end{figure}

\textit{Mutual synchronization.} In this part we assume that internally in each community the coupling is either repulsing or weakly (subcritically)
attracting and without the mutual coupling the asynchronous states $\rho_i=0$ are stable. 
To see that the mutual coupling can synchronize, we consider a simple symmetric case where the distribution widths and coupling constants are the same for all communities $\delta_i=\delta$, $\varepsilon_i=\varepsilon$, $\Gamma_i=\Gamma$, the arguments of coupling constants vanish $\alpha_i=\tilde\beta_i=0$, and there is no mismatch $\Delta=0$. Then it is easy to see from (\ref{eq:3opreal}) 
that $\Psi\to 0$. Moreover, from the  existence of a non-increasing on the trajectories of system (\ref{eq:3opreal}) Lyapunov function $L=-\Gamma\rho_1\rho_2\rho_3-\sum_{i=1}^3 (\frac{1}{2}\delta\ln(1-\rho_i^2)+\varepsilon\rho_i^2)$ it follows that in this system only equilibria are possible. According to our assumption $\delta>\varepsilon$, so the asynchronous state $\rho_1=\rho_2=\rho_3=0$ is always stable, while for large enough $\Gamma$ another synchronous state appears via a saddle-node bifurcation. In Fig.~\ref{fig:msyn} we show the regions of parameters with bistable synchrony-asynchrony states, and illustrate the appearance of the synchronous states as the mutual coupling is increased.

\begin{figure}
\centering
(a)\includegraphics[width=0.7\columnwidth]{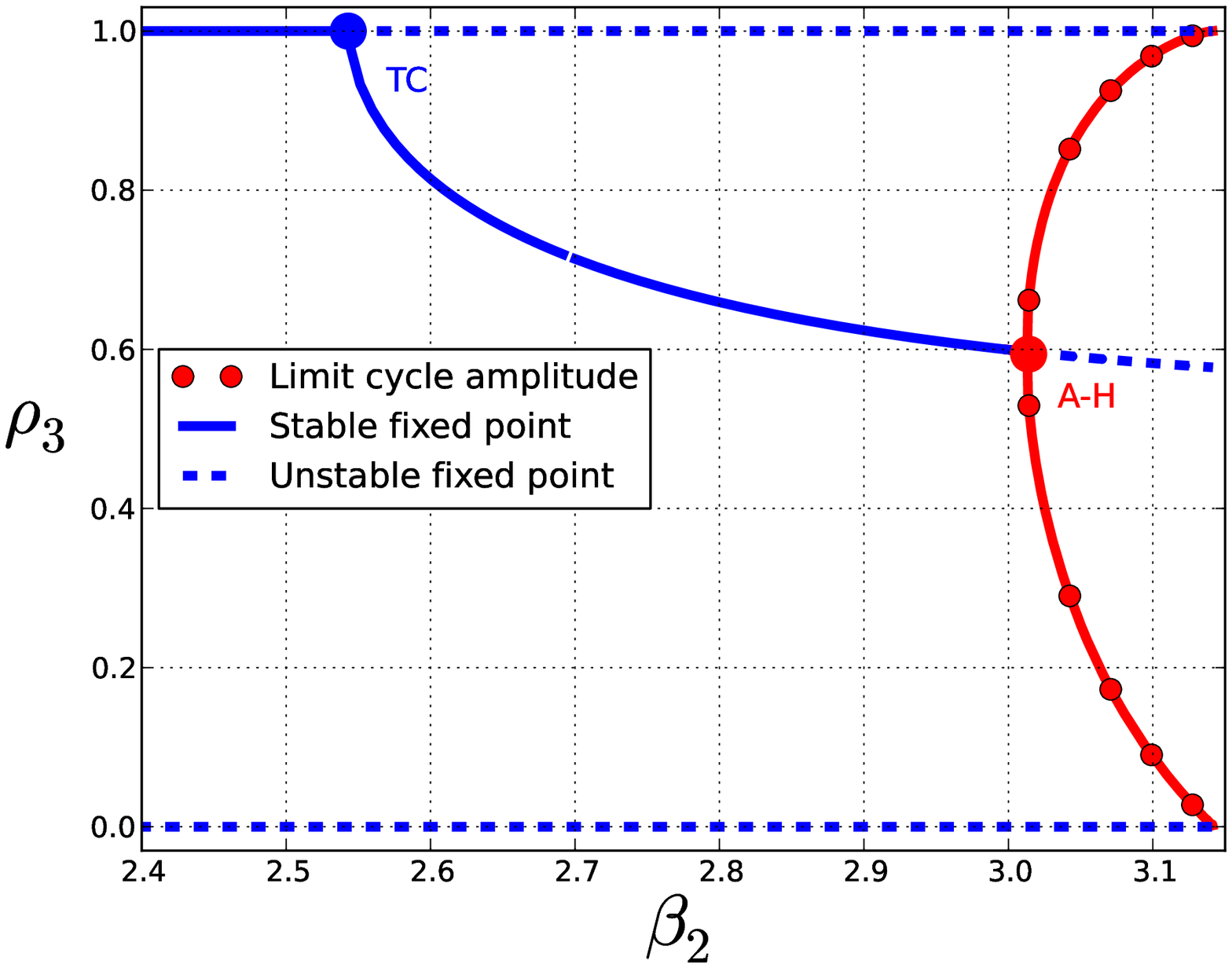}
(b)\includegraphics[width=0.7\columnwidth]{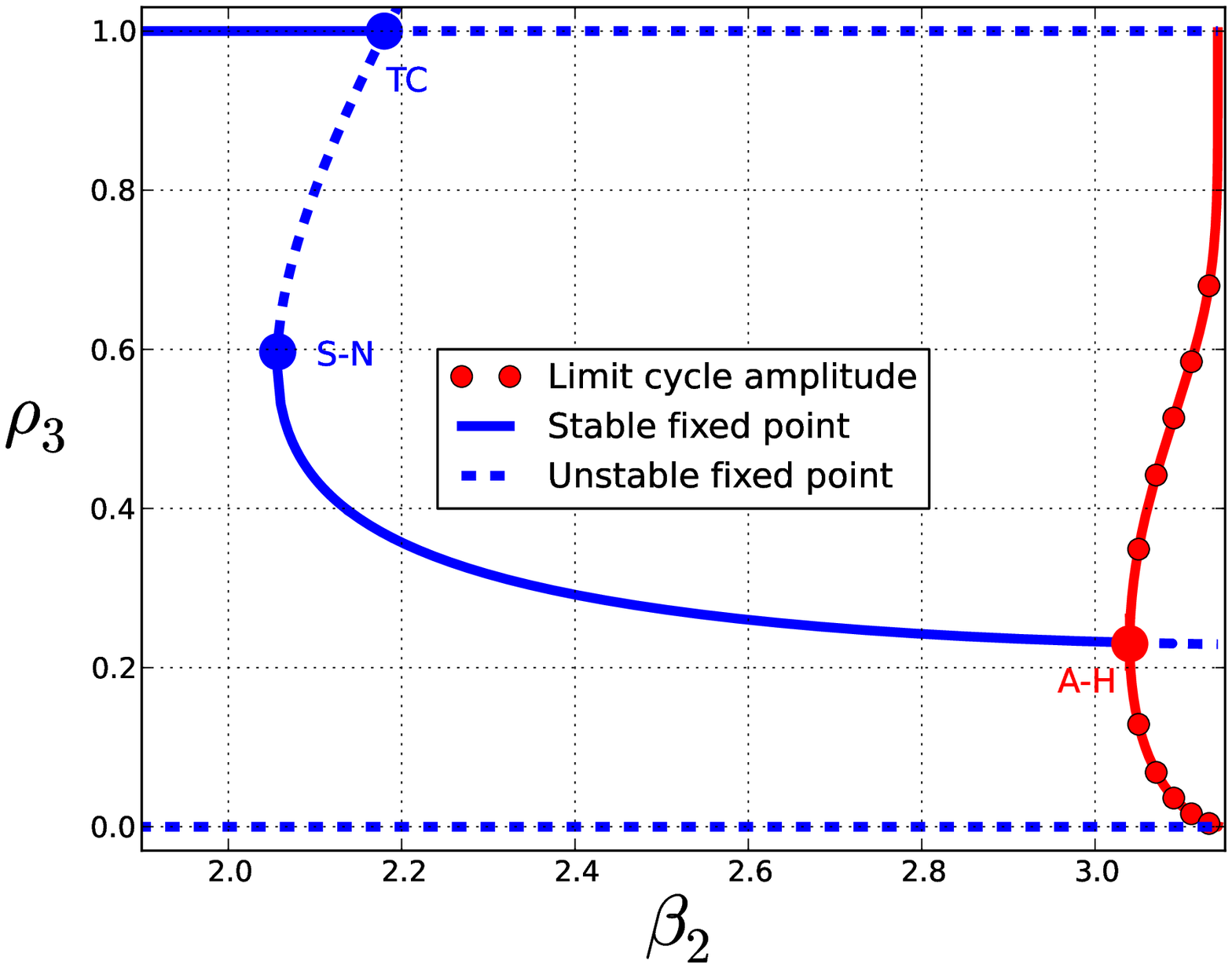}
\caption{(Color online) Bifurcation diagrams showing dependence of the order parameter $\rho_3$ on the coupling argument $\beta_2$ for $\varepsilon_i=1$, $c=2$, and different values of $b$: $b=2$ in (a) and $b=10$ in (b). Solid and dashed lines are stable and unstable equilibria; red line with markers denotes maxima and minima of periodic oscillations. TC, SN and AH denote a transcritical, a saddle-node, and an Andronov-Hopf bifurcations, respectively.}
\label{fig:desyn}
\end{figure}

\textit{Mutual desynchronization.}   As above, we assume here equal parameters for communities' heterogeneity $\delta$ and the internal coupling $\varepsilon$, so that the latter is real and exceeds the critical value for synchronization $\varepsilon>\delta$, also we set $\Delta=\alpha_i=0$. To account for possibly desynchronizing mutual interactions, we need to have non-zero arguments at least in some constants of mutual coupling.  To simplify, we assume symmetry of the two low-frequency communities $1$ and $2$, setting $\Gamma_1=\Gamma_2=b$, $\Gamma_3=c$, and $\beta_1=0$. According to Fig.~\ref{fig:stdiag}, synchronization in community $3$ may be destroyed if $\beta_2$ is close to $\pi$. 
In order to clearly see the 
desynchronization effect of the mutual coupling, we assume $\delta\to 0$, in this case the synchronous communities are in fact identical clusters with order parameters $\rho=1$, and a constant phase difference $\Psi_0=-\arctan(c\sin\beta_2(2b+c\cos\beta_2)^{-1})$. This solution is stable for small $\beta_2$, but loses stability through a transcritical bifurcation at the critical value of this parameter: $\cos\beta_{2,c}=(1-c^2-\sqrt{4bc^2-c^2+1})(2bc)^{-1}$. Beyond this transition, communities $1$ and $2$  remain in synchrony, while community $3$ becomes partially synchronized, first with a constant order parameter, and at $\beta_2$ closer to $\pi$, with a periodically oscillating one. We show the bifurcation diagram for the partial desynchronization transition in Fig.~\ref{fig:desyn}. 

\begin{figure}
\centering
(a)\includegraphics[width=0.7\columnwidth]{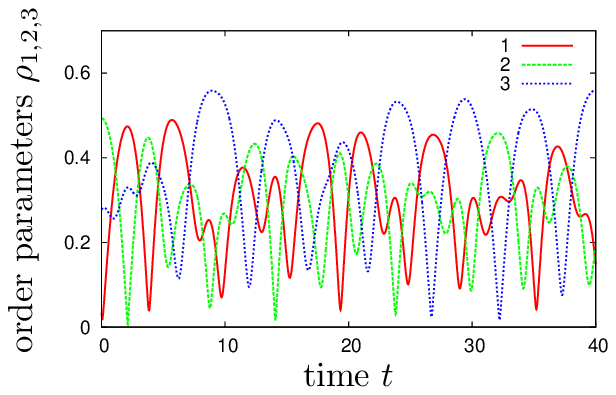}
(b)\includegraphics[width=0.7\columnwidth]{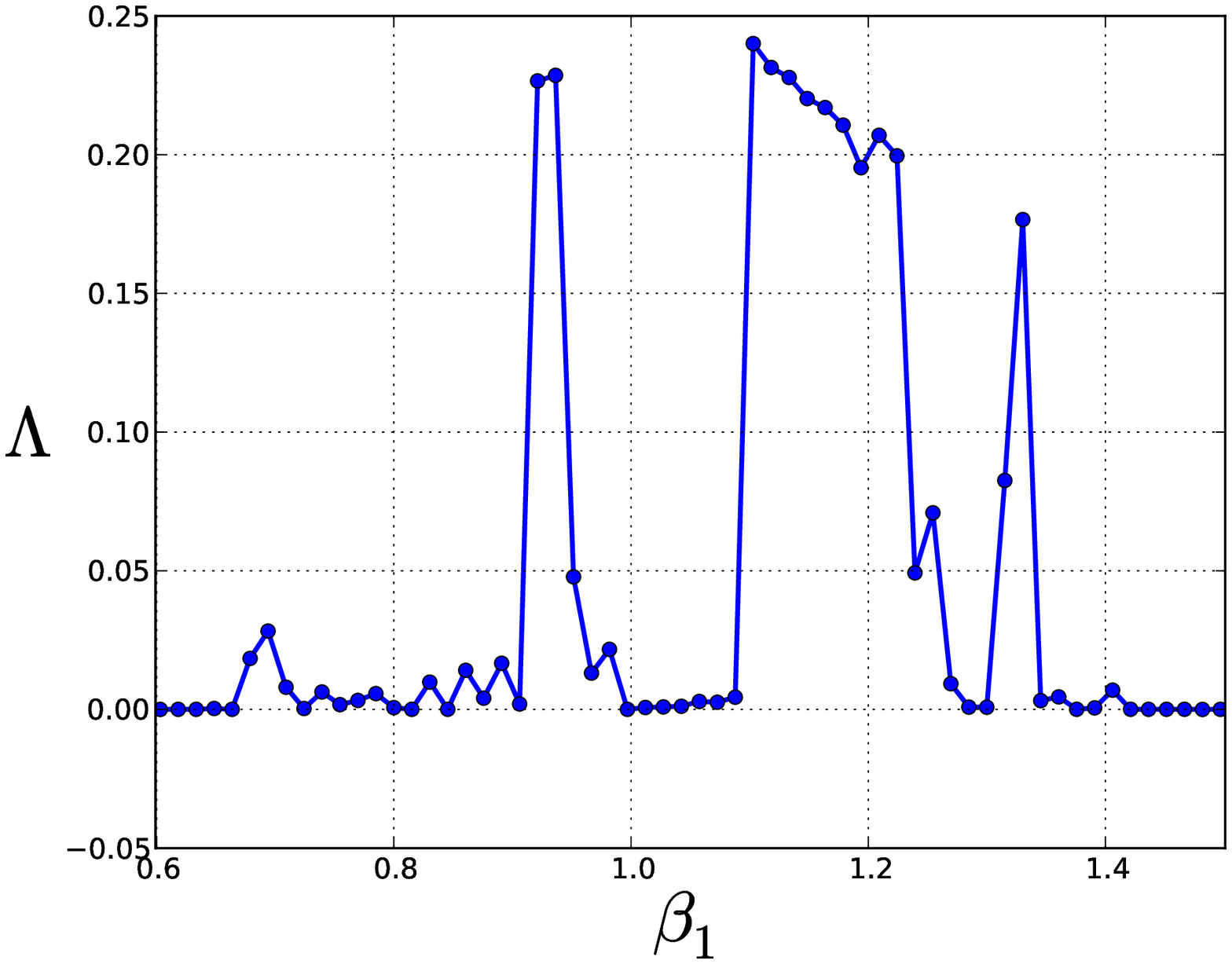}
\caption{(a) Chaotic regime in system (\ref{eq:3opreal}) for $\beta_1=1.2$. 
Time evolution of order parameters $\rho_{1,2,3}$. (b) Results of calculation of the largest Lyapunov exponent $\Lambda$ for system (\ref{eq:3opreal}) in dependence on parameter $\beta_1$. Nonzero values of $\Lambda$ indicate chaotic regime. Parameters in both cases are the following: $\delta_i=0.1$,
$\varepsilon_i=0.12$, $\alpha_i=0$, $\Gamma_i=3$, $\beta_2=\pi$, $\Delta=0.3$.}
\label{fig:chasyn}
\end{figure}

\textit{Chaotic order parameters.}  In the case when the mutual coupling between communities is much stronger than the internal one, complex synchronization patterns including chaos are possible (for other examples of chaotic order parameters in coupled communities see~\cite{Bordyugov-Pikovsky-Rosenblum-10,Kuznetsov-Pikovsky-Rosenblum-10,So-Barreto-11}). We show in Fig.~\ref{fig:chasyn} an example of such a chaotic variation of the order parameters $\rho_{1,2,3}$ 
for the case where intra-communities couplings lead to synchrony in the groups $\varepsilon>\delta$, but  due to mutual interactions chaos appears in a certain range of arguments of mutual coupling $\beta_1,\beta_2$; for real values of the
mutual coupling constants $\gamma$, i.e. for $\tilde\beta_i=0$, we have not found complex behaviors.

In conclusion, we have derived general equations describing in the phase approximation the resonant interactions between communities of oscillators, which basic frequencies differ from each other, but are in a combinational resonance.
Such a situation can be observed in populations of neurons: it is known that in brain regular macroscopic activity is observed across different frequency ranges~\cite{Buzsaki-06}. Especially alpha, gamma, and theta bands  may demonstrate quite regular oscillations, interaction of which can be treated according to the presented framework. 

We focused in this letter on a detailed description of the most elementary three-community ``triplet'' resonance $\omega_1+\omega_2\approx \omega_3$, in terms of the evolution of communities order parameters. This is accomplished by using the Ott-Antonsen ansatz allowing one to write a closed system for three complex order parameters.  Remarkably, the inter-community interaction not only shifts relative phases of the communities mean fields, but influences internal synchrony within communities. We have demonstrated how the inter-community interaction can induce or suppress internal synchronization. Furthermore, we have shown that resonant interaction
of communities can lead to chaotic dynamics of the order parameters.

M. K. thanks the G-RISC program (DAAD),  the IRTG 1740 / TRP
2011/50151-0, funded by the DFG / FAPESP, 
and the Federal Program "Scientific and scientific-educational brain-power of innovative Russia" for 2009-2013 (contract No 14.B37.21.0863) for support.


%

\end{document}